\documentclass[10pt,journal,compsoc]{IEEEtran}

\ifCLASSOPTIONcompsoc
  \usepackage[nocompress]{cite}
\else
  \usepackage{cite}
\fi

\usepackage{microtype}
\usepackage{graphicx}
\usepackage{subfigure}
\usepackage{mathtools}
\usepackage{booktabs}
\usepackage{array}
\usepackage{multirow}
\usepackage{setspace}
\usepackage{caption}
\usepackage{hyperref}
\usepackage{amsmath, amssymb, bm}
\usepackage{enumerate}
\usepackage[utf8]{inputenc}
\usepackage[english]{babel}
\usepackage{makecell}
\usepackage{amsthm}
\usepackage{color, colortbl}

\begin{document}
%
\title{FFConv: Fast Factorized Convolutional Neural Network Inference on Encrypted Data}
%
%
%
%

\author{Yuxiao~Lu, Jie~Lin, Chao~Jin, Zhe~Wang, Min~Wu, Khin~Mi~Mi~Aung, Xiaoli~Li
\IEEEcompsocitemizethanks{\IEEEcompsocthanksitem 
Yuxiao~Lu is with Singapore Management University, Singapore. E-mail: yxlu.2021@phdcs.smu.edu.sg.
Jie~Lin, Chao~Jin, Zhe~Wang, Min~Wu, Khin~Mi~Mi~Aung, Xiaoli~Li are with the Institute for Infocomm Research (I$^{2}$R), A$*$STAR, Singapore.
E-mail: \{lin-j, Jin\_Chao, wang\_zhe, wumin, mi\_mi\_aung, xlli@i2r.a-star.edu.sg\}
\protect\\
Corresponding author: Jie~Lin.

\IEEEcompsocthanksitem This work was done during Yuxiao Lu's internship at I$^{2}$R. This work was done when Lin Jie was a senior research scientist with I$^{2}$R. This work is supported by A*STAR under its AME Programmatic Funds (Project No.A19E3b0099).}

}

%
%

\markboth{In submission}%
{Yuxiao Lu \MakeLowercase{\textit{et al.}}: FFConv: Fast Factorized Convolutional Neural Network Inference on Encrypted Data}
%



\IEEEtitleabstractindextext{%
\begin{abstract}

Homomorphic Encryption (HE), allowing computations on encrypted data (ciphertext) without decrypting it first, enables secure but prohibitively slow Convolutional Neural Network (CNN) inference for privacy-preserving applications in clouds. To reduce the inference latency, one approach is to pack multiple messages into a single ciphertext in order to reduce the number of ciphertexts and support massive parallelism of Homomorphic Multiply-Accumulate (HMA) operations between ciphertexts. Despite the faster HECNN inference, the mainstream packing schemes Dense Packing (DensePack) and Convolution Packing (ConvPack) introduce expensive rotation overhead, which prolong the inference latency of HECNN for deeper and wider CNN architectures. In this paper, we propose a low-rank factorization method named FFConv dedicated to efficient ciphertext packing for reducing both the rotation overhead and HMA operations.
 FFConv approximates a $d \times d$ convolution layer with low-rank factorized convolutions, in which a $d \times d$ low-rank convolution with fewer channels is followed by a $1 \times 1$ convolution to restore the channels. The $d \times d$ low-rank convolution with DensePack leads to significantly reduced rotation operations, while the rotation overhead of $1 \times 1$ convolution with ConvPack is close to zero.
To our knowledge, FFConv is the first work that is capable of reducing the rotation overhead incurred by DensePack and ConvPack simultaneously, without introducing additional special block into the HECNN inference pipeline. Compared to prior art LoLa and Falcon, our method reduces the inference latency by up to 88\% and 21\%, respectively, with comparable accuracy on MNIST and CIFAR-10.

\end{abstract}

}

\maketitle

\IEEEdisplaynontitleabstractindextext

%
\IEEEpeerreviewmaketitle

\ifCLASSOPTIONcompsoc
\IEEEraisesectionheading{\section{Introduction}\label{sec:introduction}}
\else
\section{Introduction}
\label{sec:introduction}
\fi

Homomorphic Encryption (HE)~\cite{STOC:Gentry09,brakerski2011fully,fan2012somewhat} is one of the promising cryptographic systems that enable secure Convolutional Neural Network (CNN) inference for privacy-preserving applications in clouds while at the cost of high inference latency. At the client, the \emph{plaintext} data is encrypted in the form of \emph{ciphertext}, then transmitted to the cloud server. At the cloud, CNN inference is evaluated homomorphically on the ciphertexts to generate an encrypted prediction. The encrypted prediction is returned to the client for decryption. Since the cloud cannot encrypt or decrypt the data, the data privacy is protected. Despite the high level of security, the HE-enabled Convolutional Neural Network (HECNN) inference is prohibitively slow, mainly due to a large number of ciphertexts generated and the expensive Homomorphic Multiply-Accumulate (HMA) operations on the ciphertexts. For instance, the inference latency of a shallow CNN (1 convolution layer and 2 fully-connected layers) on one encrypted 28x28 MNIST image~\cite{MNIST} is more than 200 seconds on multi-core CPUs~\cite{MSFT:DGL+16}.

Modern HE cryptographic systems~\cite{fan2012somewhat,brakerski2014leveled,cheon2017homomorphic} used \emph{packed} encryption to accelerate the HECNN inference, in which the ciphertext structure is configured as a vector of slots, and each slot encrypts a different message (e.g., a pixel). Therefore, packing can significantly reduce the number of ciphertexts required to encrypt a given amount of data messages. In addition, packing also enables massive parallel execution of the HMA operations between ciphertexts, as per the Single-Instruction Multiple-Data (SIMD) execution model~\cite{DCC:SmaVer14}. When multiplying (resp. adding) two packed ciphertexts, it is equivalent to concurrent slot-wise multiplications (resp. additions) of the underlying vectors of the two ciphertexts. HE ciphertext packing has been employed in multiple previous works~\cite{chao2019carenets,badawi2018towards,jin2020secure} to accelerate the CNN computation and reduce resource consumption with encrypted data. In particular, LoLa~\cite{brutzkus2019low} introduced several ciphertext packing schemes, which reduced the inference latency to around 2 seconds for the shallow CNN inference on encrypted MNIST~\cite{MSFT:DGL+16}.

\begin{figure*}[ht]
\vskip 0.2in
\begin{center}
\centerline{\includegraphics[width=1.8\columnwidth]{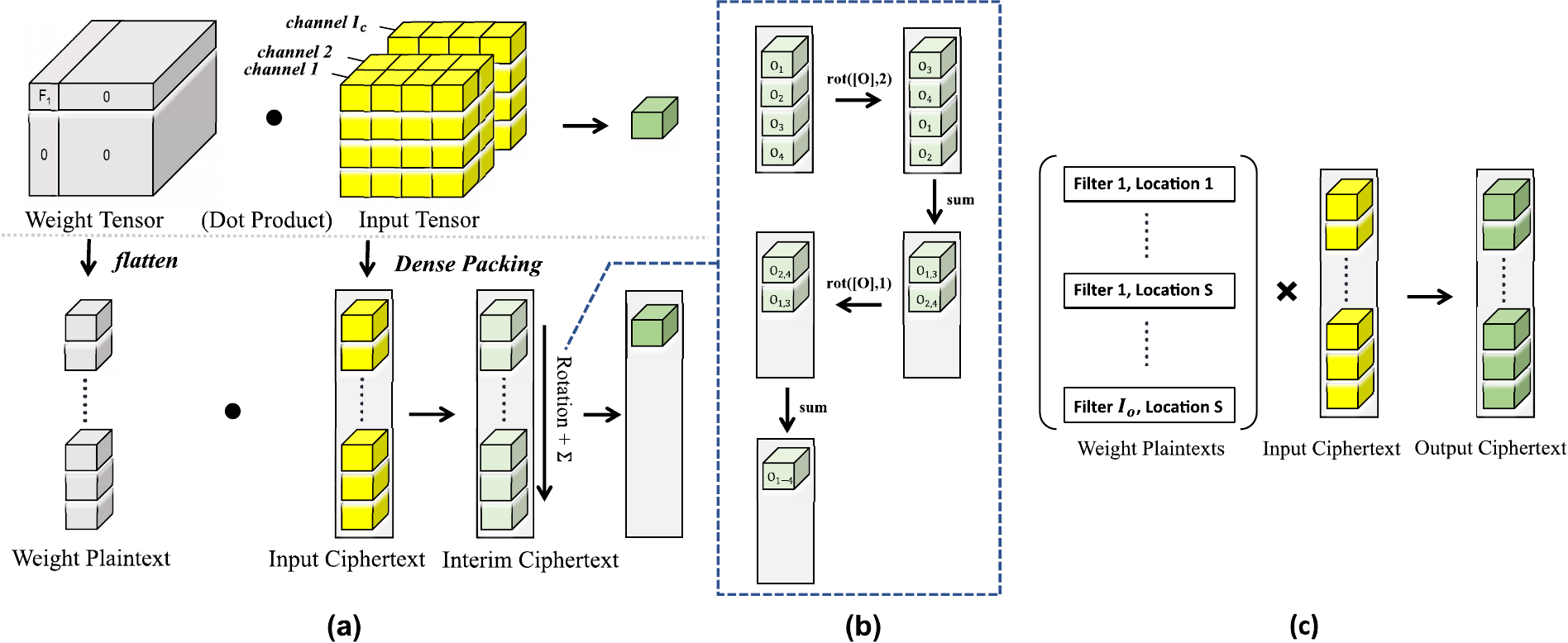}}
\caption{(a) DensePack for one filter at one shifted spatial location. (b) An example of rotation operations in homomorphic dot product to align slots in ciphertext before each addition. (c) DensePack for one convolution layer.}
\label{fig:DensePack}
\end{center}
\vskip -0.2in
\end{figure*}

\begin{figure*}[ht]
\vskip 0.2in
\begin{center}
\centerline{\includegraphics[width=1.8\columnwidth]{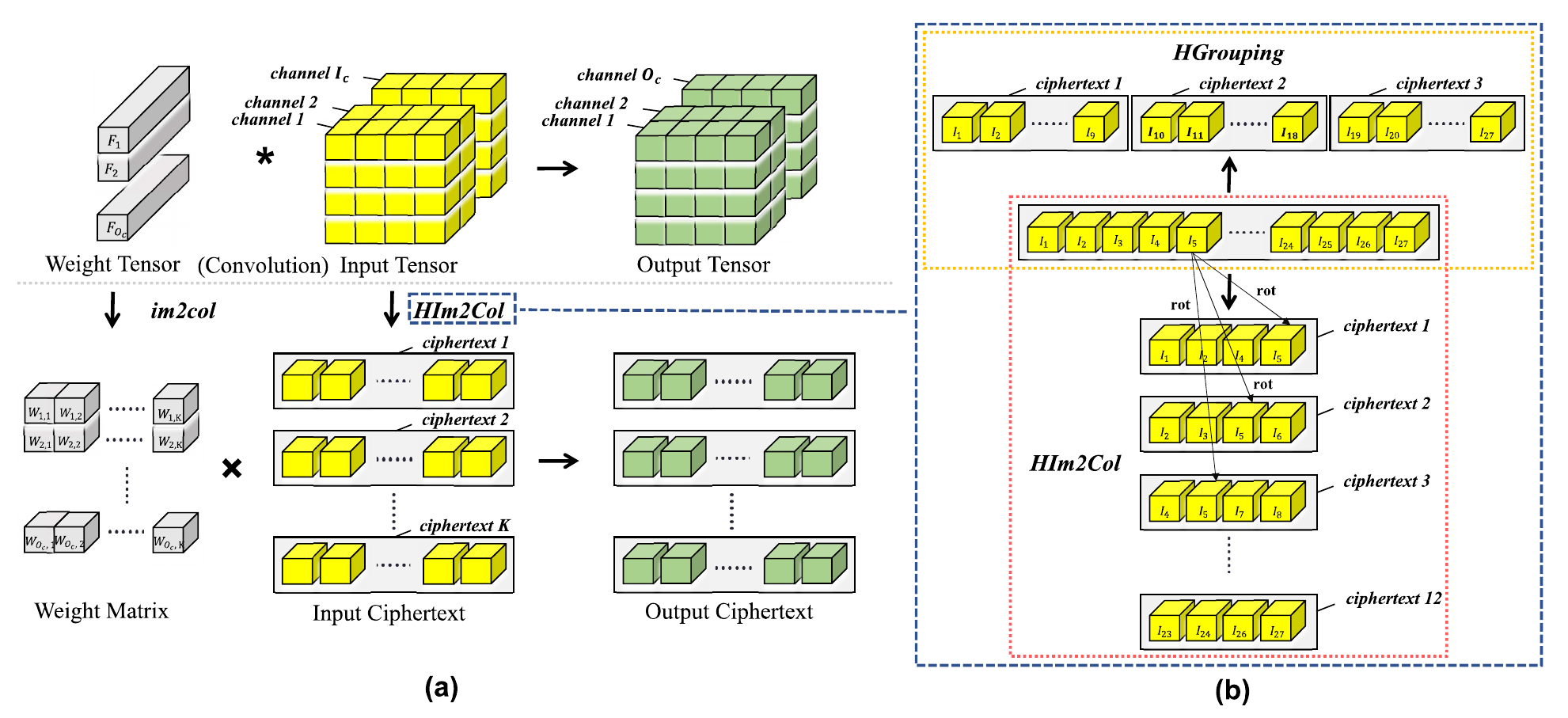}}
\caption{(a) ConvPack for one convolution layer. (b) An example of homomorphic Im2Col (HIm2Col) with convolution kernel size $d=2$ and homomorphic grouping (HGrouping) with convolution kernel size $d=1$. We assume the input 3D tensor size $I_w \times I_h \times I_c$ is $3 \times 3 \times 3$, the number of ciphertexts $K=d^2I_c$ after HIm2Col and HGrouping is 12 and 3 respectively.}
\label{fig:ConvPack}
\end{center}
\vskip -0.2in
\end{figure*}

Despite the faster HECNN inference, ciphertext packing schemes introduce expensive overhead, which prolong the inference latency of HECNN for deeper and wider CNN architectures. As summarized in LoLa~\cite{brutzkus2019low}, Dense Packing (DensePack) and Convolution Packing (ConvPack) are the two major packed representations to parallelize the HMA operations. DensePack packs 3D input tensor into a ciphertext. To execute a HMA, the \emph{Rotation} operation is required to rotate and align the slots inside the ciphertext before each accumulation operation (see Fig.~\ref{fig:DensePack} (b)). On the other side, by converting convolution operation to 2D matrix multiplication using the image-to-column (\emph{Im2Col}) technique (see Fig.~\ref{fig:lf} (top)), ConvPack packs each column of the 2D input matrix into a ciphertext, followed by the HMA operations performed over the ciphertexts without the need of any rotation operations (see Fig.~\ref{fig:ConvPack} (a)). Nevertheless, the homomorphic Im2Col operation requires a considerable number of rotation operations to reorganize the order of the slots in the input (or output) ciphertexts for the subsequent packing scheme. Compared to HMA operations, the rotations are 10x more expensive~\cite{falcon2020}. For instance, LoLa~\cite{brutzkus2019low} reported a 3-layer CNN inference (2 convolution layers and 1 fully-connected layer) on one encrypted 32x32x3 CIFAR-10 image~\cite{Krizhevsky09} takes over 700 seconds, in which the rotations account for over 90\% of the time.

Prior work attempted to further accelerate the HECNN inference by reducing the rotation overhead. Falcon~~\cite{falcon2020} replaced the convolution in spatial domain with element-wise products in frequency domain, which in turn reduced the number of rotation operations. As a result, Falcon reported 7x faster inference speed than LoLa for the 3-layer CNN inference on encrypted CIFAR-10. However, Falcon has several limitations. First, similar to LoLa, Falcon used ConvPack for the first convolution layer and DensePack for the intermediate convolution layers. There are no studies to show whether the other packing patterns (e.g. DensePack followed by ConvPack) can be integrated with CNN architectures more efficiently. Second, the frequency-domain convolution in Falcon is only applicable for reducing the rotation overhead in DensePack. How to reduce the rotation overhead introduced by homomorphic Im2Col in ConvPack remains unexplored. Third, Falcon introduced additional special blocks Homomorphic Discrete Fourier Transform (HDFT) and Homomorphic Inverse Discrete Fourier Transform (HIDFT) into HECNN, leading to more homomorphic multiplication operations.

In this paper, we propose a low-rank factorized convolution called FFConv dedicated to efficient ciphertext packing for fast inference on encrypted data. 
Our contributions are three-fold. 
\begin{itemize}

\item First, to our knowledge, this is the first work shows that the rotation overhead introduced by the homomorphic Im2Col operation can be significantly reduced under certain conditions. As shown in Table~\ref{tab:im2col}, for a convolution layer with kernel size equals 1 ($d=1$), the homomorphic Im2Col operation attached to this layer has significantly fewer number of rotations or even zero rotations, depending on the type of packing scheme of the preceding convolution layer. To take advantage of this property, one should design network architecture with $1 \times 1$ convolution layers for efficient integration with ConvPack representations, which in turn speed up the inference on encrypted data.

\item Second, we propose FFConv to approximate a $d \times d$ convolution layer with low-rank factorized convolutions, in which a $d \times d$ low-rank convolution with fewer channels is followed by a $1 \times 1$ convolution to restore the channels. Accordingly, we introduce FFConv packing which packs the factorized convolutions with either DensePack or ConvPack. The $d \times d$ low-rank convolution with DensePack leads to significantly reduced rotation operations, while the $1 \times 1$ convolution with ConvPack introduces little or zero rotation overhead for homomorphic Im2Col. To the best of our knowledge, FFConv is the first work that is able to reduce the rotation overhead incurred by DensePack and ConvPack simultaneously, without introducing additional special blocks into the HECNN inference pipeline. 

\item Lastly, compared to 8 state-of-the-art HECNNs including the most recent works LoLa and Falcon, FFConv enables efficient ciphertext packing and reduces the inference latency by up to 88\% (vs. LoLa) and 21\% (vs. Falcon), with comparable accuracy on MNIST and CIFAR-10. In addition, our method introduces significantly less noise budget than the best performing baseline Falcon. We also performed ablation studies to evaluate the effectiveness of the proposed packing pattern (versus other patterns) and the factorization method (versus training from scratch and filter pruning).

\end{itemize}

\section{Related Work}
\label{sec:rel}

\subsection{Homomorphic Encryption (HE)}

HE~\cite{STOC:Gentry09} has always been an intriguing technology due to its ability of computing on encrypted data in the absence of the decryption key. In HE, the plaintexts and ciphertexts are elements in polynomial rings. HE provides the user with two main computational operations on ciphertexts: homomorphic multiplication and homomorphic addition. These operations can manipulate ciphertexts and produce encrypted results that are equivalent to the corresponding plaintext results after decryption.

HE ciphertexts conceal plaintext messages with noise that can be identified and removed with the secret key~\cite{brakerski2011fully}. The noise magnitude can be accumulated inside a ciphertext along with the computation on it. As long as the noise is below a certain threshold that is controlled by the encryption parameters, decryption can filter out the noise and retrieve the plaintext message successfully; otherwise, the plaintext message could be corrupted and decryption could fail. Although HE schemes include a primitive (known as \textit{bootstrapping}) to reduce the noise inside ciphertexts~\cite{STOC:Gentry09}, it is extremely computationally intensive. Instead, a more practical way is to carefully select the encryption parameters to provide just enough noise budget for a ciphertext to accommodate a predefined maximum depth of computation under a specific evaluation circuit.  

In this work, following LoLa~\cite{brutzkus2019low} and Falcon~\cite{falcon2020}, we employ the Brakerski-Fan-Vercauteren (BFV) HE scheme~\cite{fan2012somewhat}. The BFV scheme is governed by three important parameters: $t$, $Q$, and $N$. First, the plaintext space is controlled by the plaintext coefficient modulus $t$. To prevent computational overflows, $t$ needs to be set to be large enough to accommodate any intermediate result of the homomorphic evaluations. Second, the scheme imposes a limit on the number of homomorphic operations that can be performed on the ciphertext before decryption fails. We refer to this limit as the computation noise budget, which can be controlled by the ciphertext coefficient modulus $Q$. Third, the underlying ring dimension $N$ is set to guarantee the targeted security level $\lambda$. For typical security requirements in practical applications, $\lambda$ is set to a minimum of $128\, \rm{bits}$. We remark that the choice of $N$ and $Q$ significantly impact the performance of HE schemes in terms of computational and memory requirements. They also affect the data expansion rate due to encryption. More specifically, the ciphertext size can be estimated to be at least $2*N*\log_{2}{Q}$ bits.

Since non-interactive HE can only support multiplication and addition operations, non-linear layers in modern neural network such as the Rectified Linear Unit (ReLU)~\cite{pmlr-v15-glorot11a} is not supported by HE. To address this issue, CryptoNets~\cite{MSFT:DGL+16} proposed to approximate ReLU with a simple Square function. There are also interactive HE solutions that combine HE with secure multi-party computation (MPC) techniques \cite{CCS:LJLA17,USENIX:JuvVaiCha18,riazi2019xonn}. These solutions use MPC to evaluate non-linear activation function, thus eliminating the need of polynomial approximation for activation function. However, they also incur high communication cost among multiple parties and require each party to have considerable computation power and be constantly online, rendering them less attractive than the non-interactive HE solutions in many use scenarios. In this paper, we focus on secure neural network inference with non-interactive HE only.

\subsection{Ciphertext Packing}

CryptoNets~\cite{MSFT:DGL+16} was the first work for secure neural network inference on encrypted data. It adopted SIMD packing which packs each pixel from a batch of images into a ciphertext, resulting in a huge number of ciphertexts and HMA operations for prediction task with batch size 1. To accelerate neural network inference with batch size 1, LoLa~\cite{brutzkus2019low} summarized another 5 ciphertext packing representations: DensePack, ConvPack, Interleaved, Stacked, and Sparse. Our work focuses on DensePack and ConvPack because they are the most often used representations for regular convolution layers, which account for over 90\% of computations in modern CNNs.
The Interleaved and Stacked representations are used for matrix multiplication in fully-connected layers, which are essentially the special case of the DensePack representation.
As such, our method can be applied to Interleaved and Stacked seamlessly.
Sparse representation is out of the scope of this paper, since the it is usually used to represent the final layer output in sparse format.

E2DM~\cite{jiang2018secure} proposed an optimized homomorphic matrix-matrix multiplication method to accelerate the homomorphic fully-connected layers. However, it requires a multiplication depth of 3 for each matrix multiplication operation, which may enlarge the HE parameters and thus lower efficiency in a deeper network scenario. Moreover, it does not provide optimizations to the homomorphic convolution layers. CHET~\cite{dathathri2019chet} and EVA~\cite{dathathri2020eva} are HE compilers with the objective of easing the burden of HE based application development. EVA optimized the low-level HE operations, while CHET and our work focus on higher-level neural network operations. Therefore, both CHET and our work can be built on top of EVA to achieve higher inference speed-up. E2DM, CHET and EVA mainly used DensePack as the ciphertext packing representation for neural network layers.
Though DensePack and ConvPack can accelerate the inference by reducing the number of ciphertexts, the packing strategies introduced expensive homomorphic rotations or Im2Col operations that prolong the inference latency of deeper and wider networks. The most recent work Falcon~\cite{falcon2020} proposed frequency-domain neural network to reduce the number of rotations in DensePack, reporting the state-of-the-art results in terms of inference time.

\subsection{Neural Networks Compression}

To build light-weight neural networks for faster inference, various compression techniques have emerged in the literature~\cite{cheng2017survey} such as quantization and pruning~\cite{zhou2016dorefa, Rastegari_2016, han2015deep_compression, ICLR2017}. Quantization represents network variables (weights and activations) with low-precision numbers, leading to smaller model size and lower computational cost. Recent works~\cite{zhou2016dorefa} mainly adopted uniform quantization to reduce the bit-width down to 2 bit without significant accuracy drop. The extreme case of quantization is binary neural networks~\cite{Rastegari_2016}, which approximate the neural network variables with 0 or 1. Quantization can been integrated with HE schemes to speed up the inference, for instance, TAPAS~\cite{pmlr-v80-sanyal18a} accelerates HE based encrypted prediction via optimizations on binary networks. Nevertheless, quantization has no dependencies on ciphertext packing.

Pruning is to remove the redundant weights from the over-parameterized neural network models, which can be categorized into unstructured weight pruning~\cite{han2015deep_compression} and structured filter pruning~\cite{ICLR2017, lin2020hrank}. The former prunes weights satisfying certain criteria such as the magnitude of weights is smaller than a pre-defined threshold. The latter prunes filters based on pruning criteria such as L1/L2 norm of filters. Faster CryptoNets~\cite{chou2018faster} used unstructured weight pruning to reduce the inference latency of HECNN by skipping the homomorphic multiplications associated with pruned weights. However, the irregular model structure after unstructured weight pruning is not compatible with ciphertext packing. On the other hand, structured filter pruning maintains regular model structure and thus can be integrated with ciphertext packing. Nevertheless, structured filter pruning often suffers from low pruning rate, which in turn limits the inference speed-up.

Another line of research in neural network compress is Low-rank factorization~\cite{Lebedev15, KimPYCYS15, Masana_2017, KossaifiTBPHP20, Idelbayev_2020_CVPR}, which decomposes a convolution layer into the product of two smaller convolutions with lower rank. Though low-rank factorization has been widely studied in accelerating neural network inference in plaintext domain, it is largely unexplored for secure neural network inference with data encrypted by HE schemes. In this work, we find a unique angle on how low-rank factorization can be integrated with ciphertext packing, in order to reduce the rotation overhead and thus achieve considerable inference acceleration on encrypted data.

\section{When Convolution meets Ciphertext Packing}
\label{sec:HECNN}
Applying HE to CNNs for private inference poses unique challenges. Typical CNNs are composed of linear and non-linear function blocks. Linear function blocks like convolution and full-connected layers, can be converted into matrix operations with simple additions and multiplications. On the contrary, non-linear function blocks usually contain complex or comparison operations such as ReLU layer~\cite{pmlr-v15-glorot11a}, which cannot be supported by non-interactive HE directly. To enable compatibility with HE primitives, there is a need to approximate these non-linear operations with polynomial functions which contain only additions and multiplications. For instance, CryptoNets~\cite{MSFT:DGL+16} suggested using the Square function to approximate ReLU in the network for classifying MNIST images.  

In a typical Machine Learning as a Service (MLaaS) scenario, network models are deployed in cloud servers to provide inference services to client users. We assume the models are kept in plaintext, while the client users encrypt their data into HE ciphertexts before sending them to the cloud server for private inference. In the next, we introduce the ways to compute convolution layers in HECNN with plaintext weights and ciphertext inputs. Pay attention to the fact that other linear layers like fully-connected and average-pooling layers can be treated as special cases of convolution layers, where for fully-connected layers, the filter sizes are the same as the input tensor sizes, and for average-pooling layers, the weights inside a single filter are set to be the same constant value.

\subsection{Convolution as Matrix Multiplication}
\label{subsec:conv}
A convolution layer is essentially dot product between filter weights and local patches cropped from the input tensor at different shifted locations. Assume the weights of a regular convolution layer are 4D tensor $\mathbf{Y} \in \mathcal{R}^{d \times d \times I_c \times O_c}$ with kernel size $d$, number of input channels $I_c$ and number of output channels $O_c$, the input and output of the convolution layer is 3D tensor $\mathbf{X} \in \mathcal{R}^{I_w \times I_h \times I_c}$ and $\mathbf{Z} \in \mathcal{R}^{O_w \times O_h \times O_c}$, where $I_w, I_h$/$O_w, O_h$ are input/output width and height respectively. Similar to the fully-connected layer, the convolution operation $\mathbf{Z}=\mathbf{X}*\mathbf{Y}$ of a convolution layer can be formulated as matrix multiplication as follows:
\begin{equation}
\mathbf{\hat{Z}} = \mathbf{I} \times \mathbf{W},
\label{eq:conv}
\end{equation}
where $\mathbf{I} \in \mathcal{R}^{S \times K}$ is a matrix with $S=O_wO_h$ rows and $K=d^{2}I_c$ columns, each row of $\mathbf{I}$ is a vector stretched out from a 3D patch $\mathcal{R}^{d \times d \times I_c}$ cropped from the input $\mathbf{X}$ for a filter at each spatial location. $\mathbf{W} \in \mathcal{R}^{K \times O_c}$ is the weight matrix, each column of $\mathbf{W}$ is a filter with $K$ parameters.

Fig.~\ref{fig:lf} (top) illustrates a regular convolution layer in the form of matrix multiplication, i.e., $\mathbf{I} \times \mathbf{W}$. Fig.~\ref{fig:lf} (bottom) illustrates $\mathbf{I} \times \mathbf{W}$ can be transformed to $\mathbf{I} \times \mathbf{W_1} \times \mathbf{W_2}$ by factorizing the weight matrix $\mathbf{W}$ as low-rank matrices $\mathbf{W_1} \times \mathbf{W_2}$.

\begin{table}[t]
\caption{Computational complexity of DensePack. Besides regular HMA operations (MulPC: plaintext-ciphertext multiplication, AddCC: ciphertext-ciphertext addition), DensePack introduces the Rotation operations. $O=O_wO_hO_c$, $O^{'}=O_wO_hO^{'}_c$.}
\label{tab:dense}
\begin{center}
\begin{sc}
\begin{tabular}{|l|c|c|c|} 
\specialrule{0em}{2pt}{2pt}
\hline
Scheme & \#MulPC & \#AddCC & \#Rot  \\ 
\hline
\specialrule{0em}{2pt}{2pt}
\hline
LoLa   & $O$        			& $Olog_{2}N$       					&  $Olog_{2}N$                           \\ 
\hline
Falcon & $\sim$3$O / p$      & $\sim$ $Olog_{2}N/p$  &  $\sim$ $Olog_{2}N/p$       \\ 
\hline
Ours   & $O'$                  & $O'log_{2}N$                         & $O'log_{2}N$       \\
\hline
\end{tabular}
\end{sc}
\end{center}
\end{table}

\begin{table}[t]
\caption{Computational complexity of ConvPack. The homomorphic Im2Col operation incurred by ConvPack is discussed in Table~\ref{tab:im2col}. $K=d^{2}I_c$.}
\label{tab:conv}
\begin{center}
\begin{sc}
\begin{tabular}{|l|c|c|} 
\specialrule{0em}{2pt}{2pt}
\hline
Scheme & \#MulPC              & \#AddCC   \\ 
\hline
\specialrule{0em}{2pt}{2pt}
\hline
LoLa   & $O_cK$     & $O_c(K-1)$      \\ 
\hline
Falcon   & $O_cK$     & $O_c(K-1)$   \\ 
\hline
Ours   & $O_c^{'}(K+O_c)$ & $O_c^{'}(K+O_c)-K-O_c^{'}$   \\
\hline
\end{tabular}
\end{sc}
\end{center}
\end{table}

\begin{table}[t]
\caption{Computational complexity of the homomorphic Im2Col (HI2C) operations when transferring the output ciphertexts of the $1^{st}$ convolution layer to the input ciphertexts of the $2^{nd}$ convolution layer. There are three possible combinations of DensePack (DP) and ConvPack (CP), for instance, DP-HI2C-CP: The $1^{st}$ convolution with DensePack followed by the $2^{nd}$ convolution with ConvPack. DP-HI2C-DP is not included in the table since homomorphic Im2Col operations are introduced by CP only. $K=d^{2}I_c$, $O=O_wO_hO_c$. ''Kernel" denotes the kernel size of the $2^{nd}$ convolution layer, with default kernel stride 1. Superscript indicates the $1^{st}$ or $2^{nd}$ convolution layer.} 
\label{tab:im2col}
\begin{center}
\begin{small}
\begin{sc}
\begin{tabular}{|c|l|c|c|c|} 
\specialrule{0em}{2pt}{2pt}
\hline
                            & Kernel        & \#MulPC        & \#AddCC             & \#Rot \\ 
\hline
\specialrule{0em}{2pt}{2pt}
\hline
\multirow{2}{*}{CP-HI2C-CP}  & $d>1$              & $O^1$      & $O^2_wO^2_hK$      &$O^2_wO^2_hK$     \\ 
	
\cline{2-5}
                            & $d = 1$ \cellcolor[gray]{.8} &  0 \cellcolor[gray]{.8}   &  0  \cellcolor[gray]{.8}   &  0 \cellcolor[gray]{.8}      \\ 
\hline
\multirow{2}{*}{DP-HI2C-CP} & $d > 1$              & $O^1$            & $O^2_wO^2_hK$      & $O^2_wO^2_hK$            \\ 
\cline{2-5}
                            & $d = 1$ \cellcolor[gray]{.8}  & $O^1_c$ \cellcolor[gray]{.8}  & 0 \cellcolor[gray]{.8} & $O^1_c$ \cellcolor[gray]{.8}  \\
\hline
\multirow{2}{*}{CP-HI2C-DP} & $d > 1$              & 0            & $I^2_c-1$              & $I^2_c-1$        \\ 
\cline{2-5}
                            & $d = 1$             & 0            & $I^2_c-1$              & $I^2_c-1$              \\ 
\hline
\end{tabular}
\end{sc}
\end{small}
\end{center}
\end{table}

\subsection{Ciphertext Packing: One Convolution}
\label{subsec:packing}

Convolution layer with packed ciphertexts has dual benefits of reduced ciphertext amount and parallelized computation. There are majorly two ways to pack the layer input into ciphertexts, namely DensePack and ConvPack, to facilitate the convolution layer to be computed in two different manners.

\textbf{Dense Packing (DensePack)}. 
As shown in Fig.~\ref{fig:DensePack} (a), for the DensePack style, the input tensor of a convolution layer is flattened as one-dimensional vector along the width, height, and channel dimensions, and then packed sequentially into a ciphertext. For one-step of convolution computation between one filter and the input ciphertext, the filter is first extended into the same size as the input tensor by padding zeros and flattened into a plaintext vector, followed by slot-wise multiplication with the input ciphertext, and then all the slots in the resultant ciphertext are accumulated to produce the convolution (dot-product) result. Fig.~\ref{fig:DensePack} (b) shows an example of the homomorphic dot product operation, one can see that rotation operation is required to align the slots before each addition/accumulation operation. Fig.~\ref{fig:DensePack} (c) shows the entire convolution layer is computed by permuting the filters at all possible shifted locations, each calculating one convolution step with the input ciphertext, and arranging all the convolution results into the final output ciphertext.

\begin{figure}[t]
\begin{center}
\centerline{\includegraphics[width=0.8\columnwidth]{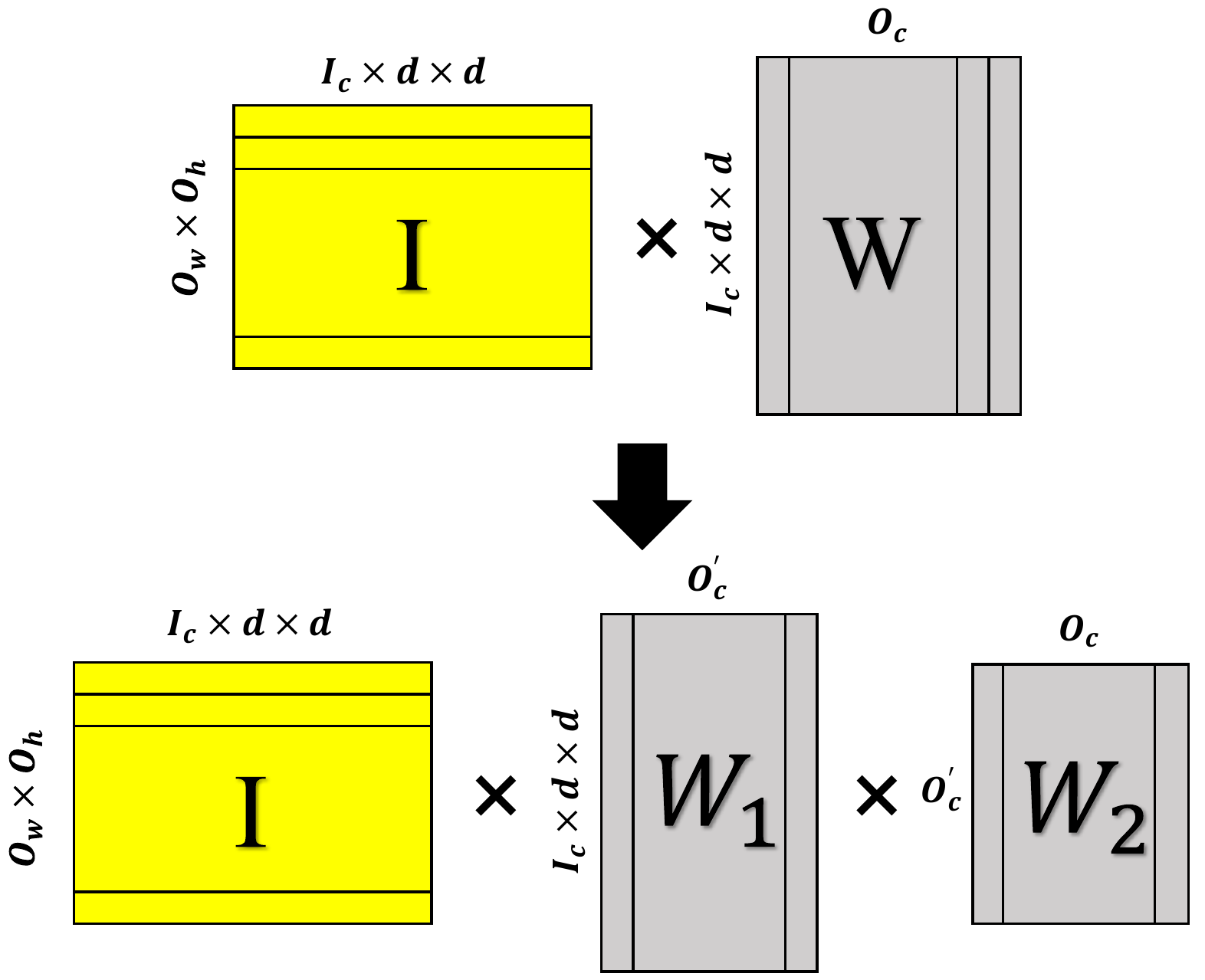}}
\caption{Low-rank factorization for a convolution layer.}
\label{fig:lf}
\end{center}
\end{figure}

\textbf{Convolution Packing (ConvPack)}.
For the ConvPack style, as shown in Fig.~\ref{fig:ConvPack} (a), the 2D input matrix of a convolution layer (see the matrix size in Fig.~\ref{fig:lf} (top)) is transformed by homomorphic Im2Col and then packed into $K$ ciphertexts, where $K=d^2I_c$ equals the number of weights in a single filter. In other works, each column of the 2D input matrix is packed into a ciphertext, which is multiplied with the filter weight. The $K$ weights in a filter are multiplied with each slot of the $K$ ciphertexts separately, and the resultant ciphertexts are added up together into one ciphertext without the need of rotation operations, which produces exactly the convolution result between the filter and the input ciphertext. Similar processes can be applied to all the $O_c$ filters in the convolution layer, and the results are $O_c$ ciphertexts each encrypts a separate channel of the output tensor.

\subsection{Ciphertext Packing: Two Convolutions}
\label{subsec:mpacking}

A typical CNN is stacked with multiple convolution layers, and it is essential to support the smooth transition of packed ciphertexts between layers, i.e., to formulate the packing of the input ciphertexts of a certain layer from the output ciphertexts of its precedent layer~\footnote{Here we only consider linear layers, as non-linear layers are computed through element/slot-wise operations, which are generally not affected by the packing schemes.}. Here we consider ciphertext packing for two consecutive convolution layers. The analysis can be directly generalized to deeper CNNs.

Fig.~\ref{fig:ConvPack} (b) illustrates an example of the homomorphic Im2Col operation, which essentially regroup the slots of input ciphertext into different locations of the new ciphertexts via rotation and HMA operations. Table~\ref{tab:im2col} shows the computational cost of homomorphic Im2Col in terms of MultPC, AddCC, and Rotation operations between two convolution layers. Generally, the overhead incurred by homomorphic Im2Col is determined by the tensor size, the kernel size, and the number of channels. Smaller kernel size usually results in smaller overhead. It must be noted that when a $1 \times 1$ kernel ($d=1$) is used, there is significantly less or even zero additional rotation overhead between two ConvPack (CP-HI2C-CP) layers or DensePack-ConvPack (DP-HI2C-CP) layers. As will be illustrated in the next section, our FFConv design takes advantage of this property to reduce the computation of convolution layers on encrypted data.

\begin{figure*}[ht]
\begin{center}
\minipage{0.5\textwidth}
  \centerline{\includegraphics[width=\linewidth]{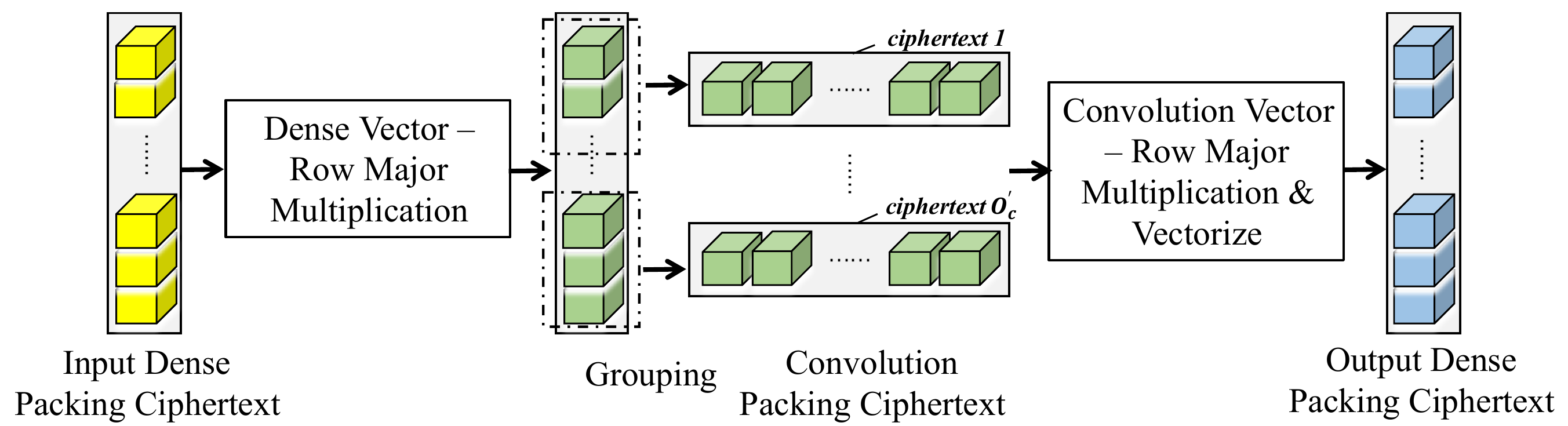}}
  \caption{DensePack-ConvPack to accelerate a regular convolution layer via low-rank factorized convolutions. For brevity, we omit the plaintext weights in the figure.}
  \label{fig:dense-conv}
\endminipage\hfill
\minipage{0.5\textwidth}%
  \centerline{\includegraphics[width=\linewidth]{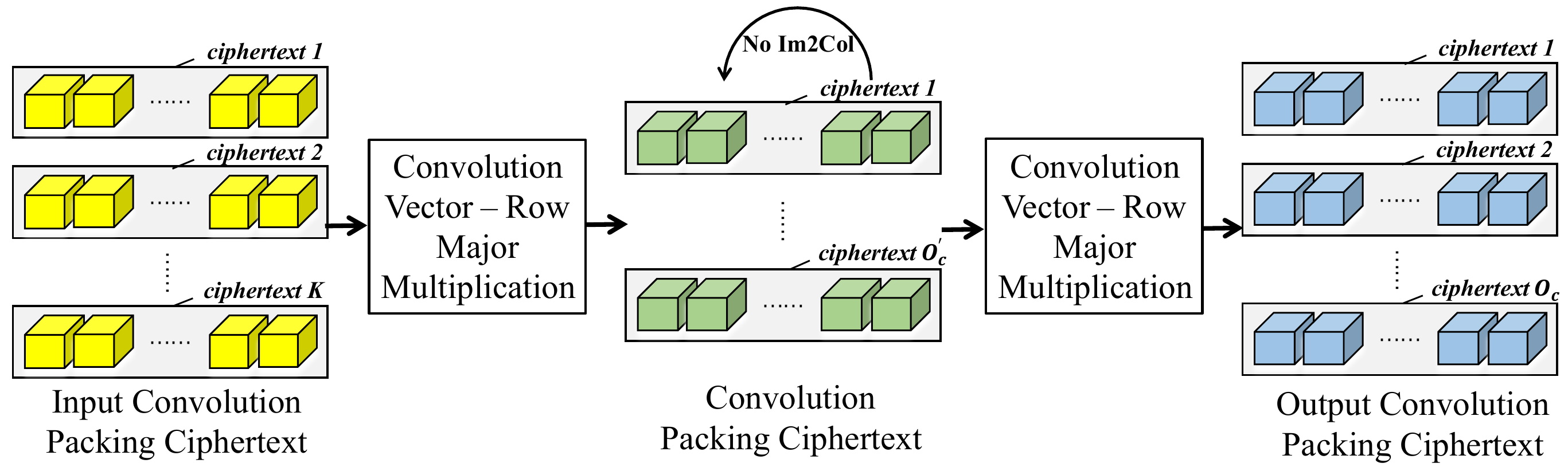}}
  \caption{ConvPack-ConvPack to accelerate a regular convolution layer via low-rank factorized convolutions. For brevity, we omit the plaintext weights in the figure.}
  \label{fig:conv-conv}
\endminipage
\end{center}
\end{figure*}

\section{FFConv}
\label{sec:ffconv}

In Section~\ref{subsec:factorization}, we introduce the low-rank factorization for convolution, which enables fast inference on data encrypted as packed ciphertexts. Section~\ref{subsec:factpack} describe how the low-rank factorized convolutions are integrated with ciphertext packing representations, which reduces the rotation overhead by a large margin and achieves significant inference speed-up compared to a regular convolution integrated with DensePack or ConvPack scheme. In Section~\ref{subsec:prior_art}, we summarize the advantages of our method over the state-of-the-art.

\subsection{Low-Rank Matrix Factorization}
\label{subsec:factorization}
Low-rank matrix factorization is a promising technique to reduce the number of multiply-accumulate operations and parameters of convolution layers~\cite{Lebedev15, KimPYCYS15}, which is achieved by factorizing the learned weight matrix $\mathbf{W} \in \mathcal{R}^{K \times O_c}$ as a product of low-dimensional matrices $\mathbf{W_1} \in \mathcal{R}^{K \times O^{'}_c}$ and $\mathbf{W_2} \in \mathcal{R}^{O^{'}_c \times O_c}$:
\begin{equation}
\begin{aligned}
\min_{\mathbf{W_1}\mathbf{W_2}} ||\mathbf{W} - \mathbf{W_1}\mathbf{W_2}||^{2}_{F} \\ 
s.t. \quad \operatorname{rank}(\mathbf{W_1}\mathbf{W_2})<O_c,
\end{aligned}
\label{eq:factor}
\end{equation}
where $||\cdot||_{F}$ is the Frobenius Norm. $\mathbf{W_1}\mathbf{W_2}$ is a low-rank approximation of $\mathbf{W}$ with the rank $O^{'}_c$ smaller than $O_c$. Based on the Eckart–Young–Mirsky theorem~\cite{eckart1936approximation}, the low-rank matrices $\mathbf{W_1}$ and $\mathbf{W_2}$ are solved analytically by the truncated Singular Value Decomposition (SVD). As a result, the number of multiplication operations is reduced from $O_cKS$ to $O^{'}_cKS$, and parameter size reduced from $O_cK$ to $O^{'}_c(K+O_c)$. As illustrated in Fig.~\ref{fig:lf} (bottom), the matrix multiplication for a regular convolution layer $\mathbf{I} \times \mathbf{W}$ is transformed to $\mathbf{I} \times \mathbf{W_1} \times \mathbf{W_2}$, with $\mathbf{W_1}$  and $\mathbf{W_2}$ are essentially equivalent to two small convolution layers. The first convolution layer is with $O^{'}_c$ $d \times d$ filters $\mathbf{W}_1 \in \mathcal{R}^{d \times d \times I_c \times O^{'}_c}$, followed by the second convolution layer with $O_c$ $1 \times 1$ filters $\mathbf{W}_2 \in \mathcal{R}^{1 \times 1 \times O^{'}_c \times O_c}$. 

The matrix factorization in Eq.~\ref{eq:factor} is data-free, only pre-trained weight matrix $\mathbf{W}$ is required to solve $\mathbf{W_1}$ and $\mathbf{W_2}$. With a pre-trained network, we apply the low-rank factorization to decompose each of the regular convolution layer (with kernel size larger than $1 \times 1$) in the network into two small convolutions with rank $O^{'}_c<O_c$. It is worth noting that the accuracy may drop with a smaller $O^{'}_c$. To restore the accuracy, one can perform re-training of the factorized network, with the weights of the two factorized convolutions initialized by the truncated SVD.

Though re-training can be used to restore the accuracy, it limits the minimum value of the rank $O^{'}_c$ since the low-rank constraint is not enforced during re-training. To further reduce the rank, one may add low rank regularization techniques such as the orthogonality constraint~\cite{Idelbayev_2020_CVPR} between $\mathbf{W_1}$ and $\mathbf{W_2}$ during re-training. As shown in our experiments, a lower rank convolution can be achieved without sacrificing the accuracy through re-training with the regularization technique, which in turn leads to further increase in inference speed-up of HECNN.

\subsubsection{Relationship to Efficient ''Bottleneck" Architecture}
In this section, we discuss the relationship of the low-rank factorized convolutions to the ''Bottleneck" architecture.
The factorized $d \times d$ convolution $\mathbf{W_1}$ with a small number of channels followed by $1 \times 1$ convolution $\mathbf{W_2}$ with a large number of channels in spirit is similar to the efficient network module \emph{Bottleneck} in the ResNet~\cite{he2016deep, heeccv}, which first reduces the number of channels with $1 \times 1$ convolution, followed by a convolution with larger kernel size ($3 \times 3$) and finally increases the number of channels with $1 \times 1$ convolution. The Bottleneck module supports end-to-end training with randomly initialized weights. Therefore, a straightforward idea is to train manually designed convolution modules $\mathbf{W_1}$ and $\mathbf{W_2}$ from scratch, without the need of low-rank factorization. However, we found that training from scratch with randomly initialized weights is inferior to low-rank factorization, which initializes the weights with truncated SVD. This is probably because the HE-enabled neural network usually used the Square function to replace the non-linear ReLU~\cite{MSFT:DGL+16}. Unlike ReLU, training with the Square function may cause instability and converge into a local minima since Square operations can easily cause the explosion of activations as network depth increases. Low-rank factorization can possibly alleviate this problem with the proper weight initialization learned from pre-trained model.

\subsubsection{Relationship to Filter Pruning}
In this section, we discuss the relationship of the low-rank factorized convolutions to filter pruning.
Filter pruning is another straightforward idea to reduce the computations of convolution layer~\cite{ICLR2017}, in which the redundant filters are identified and pruned. In this sense, filter pruning only needs to maintain the first small convolution $\mathbf{W_1}$, while low-rank factorization has two factorized convolutions. Nevertheless, low-rank factorization is superior to filter pruning. First, we observed that low-rank factorization could achieve significantly faster inference speed on encrypted data than filter pruning, with comparable accuracy. Second, as shown in Table~\ref{tab:im2col}, filter pruning has limited effect on reducing  the rotation overhead introduced by homomorphic Im2Col when ConvPack is adopted, while the factorized convolutions can.

\subsection{FFConv Packing}
\label{subsec:factpack}

Low-rank factorization decomposes a $d \times d$ convolution layer with $O_c$ filters into a small $d \times d$  convolution $\mathbf{W_1}$ with $O_c^{'}$ filters ($O_c^{'}<O_c$), followed by another $1 \times 1$ convolution $\mathbf{W_2}$ with $O_c^{'}$ input channels and $O_c$ output channels. In this section, we present FFConv Packing to pack the two convolutions $\mathbf{W_1}$ and $\mathbf{W_2}$ onto ciphertexts efficiently.

\subsubsection{ConvPack for the $1 \times 1$ Convolution $\mathbf{W_2}$} 
If we only consider the computational complexity of ConvPack itself and do not consider the homomorphic Im2Col operations incurred by ConvPack, ConvPack is much more efficient than DensePack for ciphertext packing, due to the fact that DensePack introduces a large number of expensive rotation operations (Table~\ref{tab:dense}) while ConvPack does not (Table~\ref{tab:conv}). On the other hand, as shown in Table~\ref{tab:im2col}, the computational cost of homomorphic Im2Col between the two convolution layers is significantly reduced when the kernel size of the $2^{nd}$ convolution equals 1 and ConvPack is adopted for the $2^{nd}$ convolution (i.e., DP-HI2C-CP or CP-HI2C-CP). Therefore, the most efficient packing scheme for the $2^{nd}$ $1 \times 1$ convolution $\mathbf{W_2}$ factorized by our FFConv should be ConvPack. Next, we analyze how the ConvPack for the $2^{nd}$ convolution $\mathbf{W_2}$ can be integrated with either DensePack or ConvPack for the $1^{st}$ low-rank convolution $\mathbf{W_1}$, which in turn introduces little or nearly zero rotation overheads between the two convolution layers.

\subsubsection{Packing the $d \times d$ Convolution $\mathbf{W_1}$} 
The $d \times d$ low-rank convolution layer $\mathbf{W_1}$ can be packed by either DensePack or ConvPack, depending on the layer configuration of $\mathbf{W_1}$. 

\textbf{DensePack for $\mathbf{W_1}$}. As mentioned in Section~\ref{subsec:packing}, DensePack in LoLa introduces $log_2(N) * O$ rotations into the computation process for each convolution layer, which is the main reason for the slow inference speed. Considering the inference latency incurred by DensePack is in linear relation to the number of output channels $O_c$, if we reduce $O_c$, the inference latency can be greatly saved.  As shown in Table~\ref{tab:dense}, by utilizing the property of discrete Fourier transform (DFT) with block circulant matrices, Falcon reduced the number of rotations by $p$ times ($p$ is the size of each block circulant matrix), while the value $p$ has to be a power of 2 number. Moreover, the multiplicative depth for each convolution layer in Falcon is increased from 1 to 3, which limits the network depth that can be supported. As described in Section~\ref{subsec:factorization}, another idea is to reduce $O_c$ via low-rank factorization. If we reduce $O_c$ to $O_c^{'}$ for the $1^{st}$ low-rank factorized convolution $\mathbf{W_1}$ and $O^{’}_c/O_c<= p$, we could achieve fewer number of rotations and faster inference speed than Falcon (see Table~\ref{tab:cifar}). Moreover, our method can flexibly adjust the reduction rate of inference latency according to the requirement of the model accuracy. Fig.~\ref{fig:dense-conv} illustrates an example of DensePack-ConvPack for $\mathbf{W_1}$ and $\mathbf{W_2}$.

\textbf{ConvPack for $\mathbf{W_1}$}. As shown in Table~\ref{tab:conv}, Falcon which aims to reduce rotation overhead for convolution layer with DensePack cannot be applied to reduce the inference latency of convolution with ConvPack, since there is no rotation operation incurred in ConvPack. On the contrary, our low-rank factorized convolution $\mathbf{W_1}$ with ConvPack can be accelerated because it reduces the number of channels from $O_c$ to $O_c^{'}$. As a result, the number of MulPC and AddCC operations required for ConvPack is reduced. More importantly, since the $2^{nd}$ factorized layer $\mathbf{W_2}$ is 1$\times$1 convolution with ConvPack, the homomorphic Im2Col transition of ciphertexts from $\mathbf{W_1}$ to $\mathbf{W_2}$ only involves nearly free homomorphic grouping operation, as shown in Table~\ref{tab:im2col} "CP-HI2C-CP" with $d=1$. Fig.~\ref{fig:ConvPack} (b) illustrates an example of the homomorphic grouping operation (HGrouping), which directly splits the slots of the input ciphertext into 3 ciphertexts one by one without the need of rotation. Fig.~\ref{fig:conv-conv} illustrates an example of ConvPack-ConvPack for $\mathbf{W_1}$ and $\mathbf{W_2}$.

\begin{table}[t]
\caption{Comparisons of Non-interactive HECNNs.}
\label{tab:HECNN}
\begin{center}
\begin{small}
\begin{tabular}{|l|c|c|c|c|} 
\hline
Features & CryptoNets & LoLa & Falcon & Ours  \\ 
\hline
\specialrule{0em}{2pt}{2pt}
\hline
Faster DensePack  & x    & x     & $\surd$      & $\surd$     \\ 
\hline
Faster ConvPack & x    & x    & x    &  $\surd$     \\
\hline
\end{tabular}
\end{small}
\end{center}
\vskip -0.05in
\end{table}

\subsection{Comparison with Prior Art}
\label{subsec:prior_art}

Table~\ref{tab:HECNN} summarizes the comparisons of our FFConv with state-of-the-art non-interactive HECNNs.

\textbf{Faster DensePack/ConvPack}. CryptoNets~\cite{MSFT:DGL+16} packed each pixel from a batch of images into a ciphertext, resulting in a huge number of ciphertexts as well as HMA operations for prediction at batch size 1. Though LoLa~\cite{brutzkus2019low} proposed DensePack and ConvPack to accelerate the inference by reducing the number of ciphertexts, the packing strategies introduced expensive rotation operations that prolong the inference latency of deeper and wider networks. The most recent work Falcon~\cite{falcon2020} proposed frequency-domain neural network to reduce the number of rotations for faster DensePack. However, since Falcon is not capable of handling the homomorphic Im2Col operation incurred by ConvPack, it fails to speed up the computation for convolution layer with ConvPack. On the contrary, our FFConv reduces the rotation overhead incurred by DensePack and ConvPack simultaneously. Therefore, FFConv supports a mix of packing scheme such as DensePack-ConvPack and ConvPack-ConvPack for intermediate convolution layers, leading to around 9x faster inference speed compared to the baseline packing strategy in LoLa on CIFAR-10 (see Table~\ref{tab:cifar}). Moreover, to support frequency-domain convolution in Falcon, special computational blocks HDFT and HIDFT are introduced before and after each convolution layer, which bring in 50\% more homomorphic multiplication operations compared LoLa on CIFAR-10. In contrast to Falcon, Our FFConv only factorizes a convolution layer into smaller low-rank convolutions without introducing any special components. With DensePack-ConvPack, the low-rank factorized convolutions in FFConv report 2x fewer homomorphic multiplication operations than LoLa on CIFAR-10, as shown in Table~\ref{tab:cifar}.

\textbf{Noise Budget}. Another advantage of FFConv over Falcon is that FFConv requires significantly less noise budget than Falcon. After matrix factorization, FFConv increases the multiplicative depth of a regular convolution from 1 to 2, while Falcon increases the depth from 1 to 3 due to the HDFT and HIDFT operations added to each convolution. A larger multiplicative depth would require more noise budget. As shown in Section~\ref{sec:exp}, the noise budget of our FFConv for WideNet on CIFAR-10 is 380 bits, versus Falcon 430 bits.

\section{Experiments }
\label{sec:exp}
\textbf{CNN Architectures for MNIST and CIFAR-10}.
We evaluate our method on MNIST~\cite{MSFT:DGL+16} and CIFAR-10~\cite{Krizhevsky09} datasets. MNIST contains $28 \times 28$  grayscale images divided into 60,000 training and 10,000 test samples. For MNIST, we designed a smaller neural network TinyNet, which contains only an 8*8 convolution layer with a stride of (2, 2) and 56 output channels, followed by a fully-connected layer. The accuracy of a pre-trained TinyNet with 100 epochs of training can reach 98.23$\%$. The design of TinyNet is to evaluate the effectiveness of our FFConv with ConvPack-ConvPack. Specifically, FFConv-TinyNet factorizes the first convolution layer of TinyNet into a low-rank $8 \times 8$ convolution layer (stride 2, output channels 13) and a $1 \times 1$ convolution (stride 1, output channels 56). Keeping the same training hyper-parameters, the accuracy of the factorized TinyNet can reach 98.40$\%$ with 100 epochs of re-training after the weight initialization with the truncated SVD.

CIFAR-10 is an image classification dataset containing 60,000 $32 \times 32 \times 3$ colored images for 10 object classes. The network architecture WideNet used for CIFAR-10 is the same as LoLa. The accuracy of a pre-trained WideNet with 200 epochs of training can reach 78.03$\%$. FFConv-WideNet replaces the second convolution layer of WideNet with a low-rank $6 \times 6$ convolution layer (stride 2, output channels 20) and a $1 \times 1$ convolution (stride 1, output channels 163). Accordingly, the factorized convolutions are integrated with DensePack-ConvPack for faster inference. Keeping the same training hyper-parameters, the accuracy of the factorized WideNet can reach 76.50$\%$ with 200 epochs of re-training after the weight initialization with the truncated SVD. Following LoLa~\cite{brutzkus2019low} and Falcon~\cite{falcon2020}, the weight and activations of all models for MNIST and CIFAR-10 are quantized with 8 bits.

\textbf{Cryptosystem Settings}
For fair comparisons, we use BFV scheme~\cite{fan2012somewhat} to implement all models based on the message representations and homomorphic operations used in LoLa and Falcon. We set different parameters in order to maximize the performance of each model. Specifically, (1) LoLa-TinyNet: ring dimension $N$ = 8192, plaintext coefficient modulus t = 1099511922689; (2) FFConv-TinyNet: $N$ = 8192, $t$ = 576460752303439873; (3) LoLa-WideNet: $N$ = 16384, $t$ = $34359771137 \times 34360754177$; (4) FFConv-WideNet: $N$ = 16384, $t$ = $9007199255560193 \times 9007199255658497$. We set appropriate ciphertext coefficient modulus Q respectively so that their security level is larger than 128 bits. All experiments are run on Azure standard B8ms virtual machine with 8 vCPUs and 32GB DRAM.

\begin{table}[t]
\caption{Comparison with SOTA on MNIST.}
\label{tab:mnist}
\begin{center}
\begin{tabular}{lcc}
\toprule
Method           & Time (s) & Acc (\%)  \\ 
\midrule
CryptoNets~\cite{MSFT:DGL+16}          & 205        & 98.95    \\
TAPAS~\cite{pmlr-v80-sanyal18a}       & 147         & 98.6   \\
nGraph-HE~\cite{boemer2019ngraph}           & 135        & 98.95    \\
EVA~\cite{dathathri2020eva}                 & 121.5      & 99.05    \\
FCryptoNets~\cite{chou2018faster}         & 39.1       & 98.71    \\
E2DM~\cite{jiang2018secure}                & 1.69       & 98.10    \\
LoLa~\cite{brutzkus2019low}                & 2.1        & 98.95    \\
Falcon~\cite{falcon2020}              & 1.2        & 98.95    \\
LoLa-TinyNet        & 0.45       & 98.23    \\
FFConv-TinyNet (Ours) & 0.37       & 98.40    \\
\bottomrule
\end{tabular}
\end{center}
\end{table}

\begin{table*}[t]
\caption{TinyNet with LoLa~\cite{brutzkus2019low} and our FFConv on MNIST.}
\label{tab:tinynet_mnist}
\begin{center}
\begin{small}
\begin{tabular}{|c|c|c|c|c|} 
\specialrule{0em}{2pt}{2pt}
\multicolumn{5}{c}{LoLa-TinyNet} \\
\hline
layer  & Input size & Representation & LoLa HE Operation  & Time (s)   \\ 
\hline
\multirow{2}{*}{convolution} & 64 x 144   & convolution    & convolution vector - row major multiplication          & 0.156  \\ 
\cline{2-5}
 & 54 x 144   & dense   & combine to one vector using 53 rotations and additions & 0.125  \\ 
\cline{1-5}
square  & 1 x 8064   & dense  & square   & 0.031  \\ 
\cline{1-5}
fc  & 1 x 8064   & dense & dense vector - row major multiplication  & 0.140  \\ 
\cline{1-5}
output   & 1 x 10     & dense      &  &        \\ 
\hline
 
\specialrule{0em}{2pt}{2pt}
\multicolumn{5}{c}{FFConv-TinyNet} \\
\hline
layer  & Input size & Representation & FFConv HE Operation  & Time (s)   \\ 
\hline
\multirow{3}{*}{convolution} & 64 x 144   & convolution    & convolution vector - row major multiplication  & 0.031  \\ 
\cline{2-5}
& 13 x 144   & convolution    & convolution vector - row major multiplication          & 0.046  \\ 
\cline{2-5}
 & 54 x 144   & dense          & combine to one vector using 53 rotations and additions & 0.125  \\ 
\cline{1-5}
square & 1 x 8064   & dense   & square    & 0.031  \\ 
\cline{1-5}
fc  & 1 x 8064   & dense  & dense vector - row major multiplication                & 0.140  \\ 
\cline{1-5}
output   & 1 x 10     & dense   &    &    \\
\hline
\end{tabular}
\end{small}
\end{center}
\end{table*}

\begin{table}[t]
\caption{Comparison with SOTA on CIFAR-10.}
\label{tab:cifar}
\begin{center}
\begin{small}
\begin{tabular}{lccccc} 
\specialrule{0em}{2pt}{2pt}
\toprule
   & \#MulPC & \#AddCC  & \#Rot & Time(s) & Acc(\%)  \\ 
\midrule
LoLa~\cite{brutzkus2019low}    & 8.2K & 61.0K    & 53K   & 730        & 76.5     \\
Falcon~\cite{falcon2020}  & 11.9K & 10K  & 7.9K  & 107        & 76.5     \\
Ours & 4.0K & 7.4K & 7.3K  & 84.2       & 76.5     \\
\bottomrule
\end{tabular}
\end{small}
\end{center}
\end{table}

\begin{table*}[t]
\caption{WideNet with our FFConv on CIFAR-10.}
\label{tab:widenet_cifar}
\begin{center}
\begin{small}
\begin{tabular}{|c|c|c|c|c|c|} 
\hline
Layer                        & Input size           & Representation       & FFConv HE Operation                                               & \#Rot            & Time (s)     \\ 
\hline
\specialrule{0em}{2pt}{2pt}
\hline
convolution                  & 192 x 16268 & convolution    & convolution vector - row major multiplication           & 0     & 0.65                   \\ 
\hline
square                       & 1 x 16268   & dense          & square                                                  & 0     & 0.17                   \\ 
\hline
\multirow{3}{*}{convolution} & 1 x 16268   & dense          & 500 dense vector - row major multiplication             & 7000  & 73.45                  \\ 
\cline{2-6}
                             & 20 x 25     & convolution    & convolution vector - row major multiplication           & 0     & 1.82                   \\ 
\cline{2-6}
                             & 163 x 25    & dense          & combine to one vector using 162 rotations and additions & 162   & 7.97                   \\ 
\hline
square                       & 1 x 4075    & dense          & square                                                  & 0     & 0.17                   \\ 
\hline
fc                           & 1 x 4075    & dense          & 10 dense vector - row major multiplication              & 120   & 1.67                   \\ 
\hline
output                       & 1 x 10      & dense          &                                                         &       & \multicolumn{1}{l|}{}  \\
\hline
\end{tabular}
\end{small}
\end{center}
\end{table*}

\subsection{Comparison with SOTA on MNIST}
Table~\ref{tab:mnist} shows the comparisons in terms of inference latency and accuracy between our FFConv-TinyNet and the state-of-the-art on MNIST. Compared to the baseline LoLa-TinyNet, FFConv-TinyNet accelerates the inference speed by 17.78$\%$, from 0.45 seconds to 0.37 seconds. With comparable accuracy, FFConv-TinyNet is significantly faster than the SOTA including nGraph-HE~\cite{boemer2019ngraph}, Faster CryptoNets~\cite{chou2018faster}, E2DM~\cite{jiang2018secure}, LoLa~\cite{brutzkus2019low}, and Falcon~\cite{falcon2020}. The most recent work Falcon reported inference latency 1.2 seconds with accuracy 98.95\%, while our method is 0.37 seconds with accuracy 98.4\%. The 3x speed-up is mainly attributed to fact that the FFConv-TinyNet with ConvPack-ConvPack strategy is rotation free, while Falcon still has rotation operations even with the optimized frequency-domain convolution. It is worth noting that the introduction of FFConv has slightly improved the accuracy of original TinyNet. The possible reason is FFConv also reduces the amount of model parameters, which can help prevent model overfitting and enhance generalization ability.

Table~\ref{tab:tinynet_mnist} summarizes the message representation, homomorphic operation and inference latency that LoLa and FFConv applied at each layer of TinyNet. Both LoLa-TinyNet and FFConv-TinyNet implemented plaintext Im2Col to preprocess the input and encode the input into 64 ciphertexts, with each ciphertext contains 144 elements. After performing convolution vector-row major multiplication on each of the 64 ciphertexts, LoLa-TinyNet and FFConv-TinyNet generate 54 dense output messages and 13 dense output messages and consume 0.156 seconds and 0.031 seconds, respectively. For FFConv-TinyNet, an additional layer of convolution vector-row major multiplication is required to form the entire first convolution layer, which results in dense output messages in 0.046 seconds. Although FFConv-TinyNet uses two layers of convolution vector-row major multiplication, it still reduces the first convolution layer inference time from 0.156 seconds by 50.64$\%$ to 0.077 ($0.031 + 0.046$) seconds. The remaining layers of FFConv-TinyNet are the same as LoLa-TinyNet and therefore use the same time.

\subsection{Comparison with SOTA on CIFAR-10}
Table~\ref{tab:cifar} shows the comparisons in terms of HMA and rotation operations, inference latency and accuracy between our FFConv and the state-of-the-art with WideNet on CIFAR-10. The inference latency of FFConv is 84.2 seconds, which reduces the inference latency of LoLa and Falcon by 88\% and 21\% respectively. Table~\ref{tab:widenet_cifar} summaries the message representation, homomorphic operation and the number of rotations that FFConv applies at each layer of WideNet. In LoLa, the second convolution layer of WideNet is with DensePack and consumes 711 seconds, accounting for more than 97$\%$ of the total inference latency. This is mainly due to the large number of time-consuming rotation operations in DensePack. Therefore, we focus on optimizing the second convolution layer of WideNet, which contains nearly 500,000 parameters. FFConv replaces this layer with two factorized convolution layers, which reduce the number of rotations by 86.48$\%$ (from 52,975 to 7000 + 162 = 7162), the number of MulPC operations by 7.75$\%$ (from 4075 to 3575), and the number of AddCC operations by 81.61$\%$ (from 52,975 to 9740). The message representation and homomorphic operation of the remaining layers remain unchanged, so the execution time are the same as LoLa. On the other hand, Falcon transformed the second spatial-domain convolution layer of WideNet into frequency-domain convolution in order to reduce the number of rotations. However, Falcon introduced a large amount of MulPC ($\sim$3.7K) into the inference pipeline. Compared to Falcon, the number of rotation operations in FFConv is 9.07$\%$ less. Moreover, the increase in noise caused by Falcon's HDFT and HIDFT operations requires higher noise budget $Q=430$-bit, while FFConv-WideNet only needs 380-bit.

\begin{table}[t]
\caption{Execution time of the FFConv Packing variants for the two factorized convolutions $\mathbf{W_1}$ and $\mathbf{W_2}$ in WideNet on CIFAR-10. CP: ConvPack, DP: DensePack, HI2C: Homomorphic Im2Col between $\mathbf{W_1}$ and $\mathbf{W_2}$.}
\label{tab:packing-conv-variants}
\centering
\begin{tabular}{|c|c|c|c|c|} 
\hline
                    & $\mathbf{W_1}$ (s) & $\mathbf{W_2}$ (s) & HI2C (s) & Total (s)  \\ 
\hline
CP-HI2C-DP  & 28.65          & 711            & 2530            & 3269.65         \\
\hline
CP-HI2C-CP   & 28.67          & 1.82           & 2530            & 2560.49         \\ 
\hline
DP-DP & 73.45          & 711            & 0               & 784.45          \\ 
\hline
DP-HI2C-CP  & 73.45          & 1.82           & 0               & 75.27           \\ 

\hline
\end{tabular}
\end{table}

\begin{table}[t]
\caption{Low-rank factorized convolutions with truncated SVD initialization or random initialization during the re-training of FFConv-WideNet on CIFAR-10. We set the rank $O_c^{'}$ as \{15, 20, 25\}.}
\label{tab:init_rank}
\centering
\begin{tabular}{ c | c  c  c  c } 
\hline
\multirow{2}{*}{$O_c^{'}$} & \multicolumn{2}{c}{SVD-init}   & \multicolumn{2}{c}{Random-init}                        \\
                  & Mean & Var & Mean & Var \\ 
\hline
\specialrule{0em}{2pt}{2pt}
\hline
15                                       & 74.874                   & 0.02173                 & 74.166                   & 0.06463                  \\
20                                       & 76.672                   & 0.02972                 & 75.298                   & 0.06067                  \\
25                                       & 77.270                    & 0.03275                 & 76.232                   & 0.01547                  \\
\hline
\end{tabular}
\end{table}

\begin{table}[t]
\caption{Comparison of re-training stability when initializing the factorized convolutions with truncated SVD (SVD-Init) or Random (Rand-init), with deeper network DeepNet~\cite{CCS:LJLA17} trained on CIFAR-10.}
\label{tab:init_lr}
\begin{center}
\begin{tabular}{c|cccc} 
\hline
\multirow{2}{*}{lr} & \multicolumn{2}{c}{SVD-Init} & \multicolumn{2}{c}{Random-Init}  \\
                    & Acc (\%)     & Test Loss               & Acc (\%)     & Test Loss            \\ 
\hline
\specialrule{0em}{2pt}{2pt}
\hline
10\^{}-4            & 81.74 & 0.62             & 10.00 & 2.30          \\
10\^{}-5            & 75.97 & 0.74             & 10.00 & 2.30          \\
\hline
\end{tabular}
\end{center}
\end{table}

\subsection{Ablation Study}

\subsubsection{Variants of FFConv Packing}

To verify the effectiveness of the FFConv packing schemes chosen for our method, we evaluate the inference time of all possible variants of the packing combinations with DensePack and ConvPack. We choose the low-rank factorized convolutions for the second convolution layer of WideNet trained on CIFAR-10, which accounts for more than 90\% of the inference time for WideNet. For the factorized low-rank $6 \times 6$ convolution layer with output channels 20 and the $1 \times 1$ convolution layer with output channels 163, there are in total 4 possible combinations of DensePack and ConvPack, namely, ConvPack-DensePack (CP-HI2C-DP), ConvPack-ConvPack (CP-HI2C-CP), DensePack-DensePack (DP-DP) and DensePack-ConvPack (DP-HI2C-CP). Table~\ref{tab:packing-conv-variants} shows the total inference time and the breakdown for the 4 packing combinations. One can see that DP-HI2C-CP used for our FFConv-WideNet is the best performing one, which is 10x faster than the DP-DP scheme adopted by LoLa~\cite{brutzkus2019low}. The trend is consistent with the computational complexity analysis in Section~\ref{sec:HECNN} (see Table~\ref{tab:dense} to Table~\ref{tab:im2col}). In Table~\ref{tab:im2col} and Table~\ref{tab:packing-conv-variants}, it is worth noting that CP-HI2C-CP is ~34 times slower than DP-HI2C-CP. Though CP-HI2C-CP does not introduce rotation overhead in the homomorphic Im2Col operation between the factorized convolution layers, the first $3 \times 3$ factorized low-rank convolution incurs a large number of rotation operations when transforming the output ciphertext from the preceding convolution layer through homomorphic Im2Col, as indicated in Table~\ref{tab:im2col} ($d>1$).

\subsubsection{Why Low-rank Factorization?}

The low-rank factorization ($\mathbf{W_1}$ and $\mathbf{W_2}$) in FFConv in spirit is similar to the manually designed efficient network module \emph{Bottleneck} in ResNet~\cite{heeccv}. The Bottleneck module supports end-to-end training with randomly initialized weights. Therefore, one may choose to train the convolution modules $\mathbf{W_1}$ and $\mathbf{W_2}$ with random initialization, without the need of low-rank factorization. However, we find that training $\mathbf{W_1}$ and $\mathbf{W_2}$ from scratch with randomly initialized weights (Random-Init) is inferior to low-rank factorization, which initializes the weights with the truncated SVD (SVD-Init).

First, we evaluate the effect of weight initialization with WideNet on CIFAR-10. We set $O_c^{'} = 15, 20, 25$, and train the factorized network with weights initialized by either random initialization or the truncated SVD initialization. A pre-trained WideNet reference model with accuracy 78.03\% is used to calculate the truncated SVD for weight initialization. For training with both SVD-Init and Random-Init, the initial learning rate is lr = $10^{-3}$ and multiplied by 0.97 after each epoch. We use Adam as the optimizer and train each model for 100 epochs. For each model, we repeat the training for 5 times with random seeds and report the averaged accuracy with variance. As shown in Table~\ref{tab:init_rank}, we observe that SVD-Init performs consistently better than Rand-Init at different $O_c^{'}$.

Second, we further increase network depth to explore the effect of weight initialization on the training stability and convergence for deeper networks. We choose DeepNet~\cite{CCS:LJLA17}, a 7-layer convolutional neural network which contains 6 convolution layers, 6 Square activation functions and 1 fully connected layer. We apply factorization to the convolution layers except for the first layer, i.e., the number of kept filters is $\{7, 13, 13, 25, 25\}$ from the $2^{nd}$ to the $6^{th}$ convolution layer. We train a DeepNet reference model achieving 86\% accuracy on CIFAR-10 for calculating the truncated SVD. For re-training with SVD-Init and Random-Init, we initialize the learning rate (lr) with values from $10^{-4}$ to $10^{-5}$, and lr is multiplied by 0.98 after each epoch during training. We train each model for 200 epochs. As shown in Table~\ref{tab:init_lr}, one can see that the training with SVD-Init achieves accuracy at 81.74\% with initial lr as $10^{-4}$, while the training with Random-Init failed to converge regardless of the initial lr used.

Since the activation function used in both WideNet and DeepNet is the Square function~\cite{MSFT:DGL+16}, its partial derivative could be larger or even exploded during back-propagation. Compared to training with popular activation functions such as ReLU, training with the Square function is easier to cause instability and converge into a local minima. Therefore, it is helpful to avoid the local minima through proper initialization of the weights before training, suggesting that the low-rank factorization is important for training deep networks with Square activation functions in the context of fast and secure neural network inference on encrypted data.

\begin{figure}[t] 
\centering 
\includegraphics[width=0.5\textwidth]{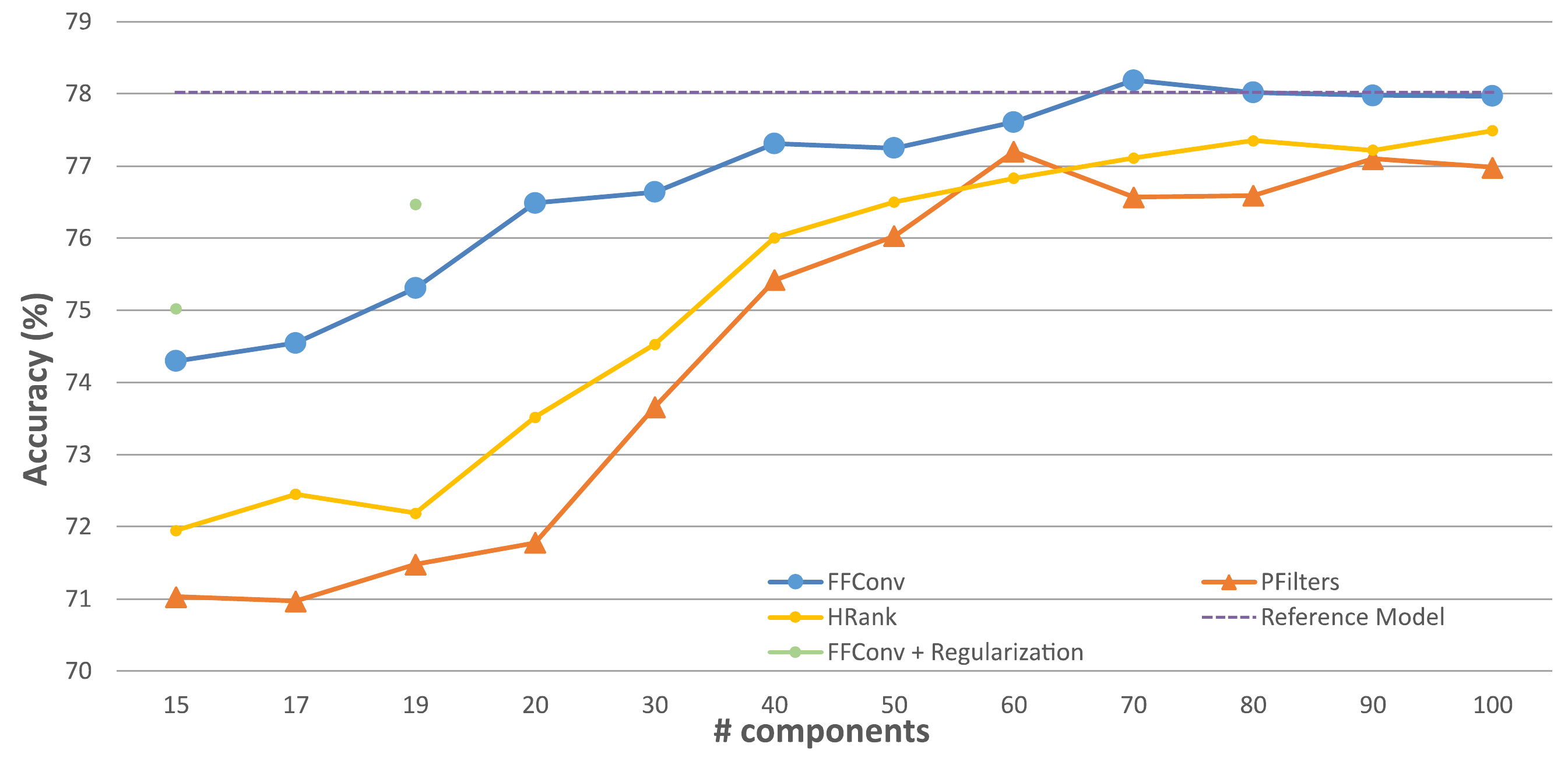}
\caption{Comparisons of FFConv (w or w/o low-rank regularization during re-training) with state-of-the-art filter pruning approaches PFilters~\cite{ICLR2017} and HRank~\cite{lin2020hrank}.} 
\label{fig:fpff} 
\end{figure}

\subsubsection{FFConv versus Filter Pruning}

Besides FFConv, filter pruning~\cite{ICLR2017, lin2020hrank} can also speed up the inference of convolution layer on encrypted data, by reducing the number of filters from $O_c$ to $O^{'}_c$. Nevertheless, we observe that FFConv achieves a higher compression rate than filter pruning at comparable accuracy. As a result, the inference speed of FFConv on encrypted data is significantly faster than that of filter pruning. 

We evaluate FFConv and filter pruning with WideNet~\cite{falcon2020} on CIFAR-10. Since the second convolution layer of WideNet is with $O_c = 163$ filters and accounts for more than 90\% of the total inference latency, we optimize this convolution layer via FFConv or filter pruning. A pre-trained WideNet reference model achieves accuracy 78.03\%, which is used for the subsequent low-rank factorization or filter pruning. For a fair comparison, we set the same hyper-parameters for both FFConv and filter pruning during retraining. The initial learning rate is lr = $10^{-3}$ and multiplied by 0.97 after each epoch. We use Adam as the optimizer. For filter pruning, we use the L1 norm based automated gradual pruning algorithm PFilters~\cite{ICLR2017, ZhuG18} or the most recent work HRank~\cite{lin2020hrank} as the pruning scheduler. For PFilters~\cite{ICLR2017}, the pruning rate is progressively increased every two epochs in the first 30 epochs till it reaches the target pruning rate. Both FFConv and filter pruning models are re-trained for 100 epochs. 

Fig.~\ref{fig:fpff} shows the results of FFConv and filter pruning approaches as a function of pruning rate (i.e. \# components $O_c^{'}$ kept). When $O_c^{'} > 70$, FFConv performs on par with the reference model, while filter pruning performs slightly worse by around 1\%. As $O_c^{'}$ is further reduced, the accuracy of FFConv drops slowly, while the accuracy gap between FFConv and filter pruning approaches becomes larger, especially at low $O_c^{'}$. For instance, if the number of kept components $O_c^{'}$ is less than 20, FFConv outperforms HRank by over 3\%. This is probably because FFConv is able to restore the model's capacity as much as possible, through the $2^{nd}$ factorized 1$\times$1 convolution.

The $2^{nd}$ factorized convolution by FFConv introduces little overhead, which is negligible compared to the $1^{st}$ factorized convolution which dominates the computations on encrypted data especially when $O_c^{'}$ is large. As shown in Fig.~\ref{fig:fpff}, we observe that FFConv achieves significantly smaller $O_c^{'}$ than filter pruning at comparable accuracy, which in turn reduces the inference latency on encrypted data by a large margin. When $O_c^{'} = 20$, the accuracy of FFConv-WideNet is 76.5\%, with the inference latency 84.2 seconds. At comparable accuracy, filter pruning has to keep $O_c^{'}=50$ to $O_c^{'}=60$ filters, with inference latency at 204 seconds and 241 seconds respectively, which is more than 2x slower than FFConv.

\subsubsection{Re-training FFConv with Low-rank Regularization}
As shown in the previous section, re-training FFConv with weights initialized by truncated SVD of a pre-trained model can restore accuracy by carefully choosing the rank $O^{'}_c$ of the factorized convolutions $\mathbf{W_1}$ and $\mathbf{W_2}$. In this section, we show that one can further reduce the rank without incurring accuracy loss by adding low-rank regularization during re-training, which in turn accelerate the inference of HECNNs. For simplicity, we follow the ''Learning-Compression" optimization in~\cite{Idelbayev_2020_CVPR}, and enforce the orthogonality constraint between $\mathbf{W_1}$ and $\mathbf{W_2}$ during re-training while decreasing the rank manually at each optimization step. As shown in Fig.~\ref{fig:fpff}, FFConv with regularization during re-training outperforms the vanilla FFConv by $0.72\%$ and $1.16\%$ with $O^{'}_{c} = 15$ and $O^{'}_{c} = 19$, respectively.

\section{Conclusion}
\label{sec:con}
In this paper, we propose a low-rank factorization approach named FFConv to accelerate secure neural network inference on encrypted data with ciphertext packing. FFConv factorizes a regular convolution into two low-rank convolutions, in which the input ciphertexts can be packed with DensePack and ConvPack efficiently. Experimental results show that FFConv enables faster inference speed than state-of-the-art. FFConv is the first non-interactive HECNN that is capable of reducing the rotation overheads incurred by DensePack and ConvPack packing schemes simultaneously.

\ifCLASSOPTIONcaptionsoff
  \newpage
\fi



%

\bibliography{ffconv}
\bibliographystyle{IEEEtran}


%




\end{document}